\documentclass[12pt]{iopart}
\usepackage{graphicx}
\usepackage{amssymb}
\begin{document}

\title{$J/\psi$ production and elliptic flow parameter $v_2$ at LHC energy}

\author{R. Peng$^{1,2,3}$ and C. B. Yang$^{1,2}$}

\address{$^{1}$Institute of Particle Physics, Hua-Zhong Normal
University, Wuhan 430079, China\\
$^{2}$ Key Laboratory of Quark and Lepton (Hua-Zhong Normal University),
Ministry of Education, China\\
$^3$College of Science, Wuhan University of Science and Technology, Wuhan 430065, China}
\ead{pengru$\_$1204@hotmail.com}
\begin{abstract}
We apply the recombination model to study $J/\psi$ production and its elliptic flow in the region of $10<p_T<20$ GeV/$c$ at LHC energy. We show the distribution of $J/\psi$ as function of the transverse momentum $p_T$ and the azimuthal angle $\phi$. If the contribution from the recombination of shower partons from two neighboring jets can not be ignored due to the high jet density at LHC, the elliptic flow parameter $v_2$ of $J/\psi$ is predicted to decrease with $p_T$.
\end{abstract}

\vspace{2pc}
\maketitle

In the framework of the recombination model, the mechanism of hadronization is described as recombinations of
thermal-thermal partons, thermal-shower partons and shower-shower partons\cite{CBYpion}. With the shower parton distributions (SPD) obtained from fitting the fragmentation functions (FF) \cite{CBYShower,pengru1}, the model has given good agreement with the experimental data of the hadron production at relativistic heavy ion collisions (RHIC) \cite{CBYpion}-\cite{strange}. In Pb+Pb collisions at $\sqrt{s_{NN}}=5.5$ TeV at Large Hadron Collider (LHC), the initial temperature and energy density higher than those at RHIC offer a new domain to study the physics of strongly interacting matter and the properties of quark gluon plasma (QGP). The recombination model has also been applied to a large $p_T$ range of $10<p_T<20$ GeV/c at LHC and predicted a new phenomenon that the proton-to-pion ratio can be as high as 20, much higher than that at RHIC \cite{CBYLHC}. Charm quark, abundantly produced at LHC is one of the main observables which can trace the initial phase of the collision and provide information on the possible formation of QGP. For this reason, we study $J/\psi$ production in central Pb+Pb collisions (0-10\% centrality) at LHC in this paper.

In the recombination model the meson production is expressed as the sum of $\mathcal{TT}$ (thermal-thermal), $\mathcal{TS}$ (thermal-shower) and $\mathcal{SS}$ (shower-shower) terms, since there are two components of partons, thermal ($\mathcal{T}$) partons and shower ($\mathcal{S}$) partons originated from hard partons produced in hard partonic interactions. While for $\mathcal{SS}$ term, we should consider the shower partons from one jet and shower partons from two neighboring different jets and we denote the one-jet contribution as $\mathcal{SS}(1)$ and the two-jet $\mathcal{SS}(2)$. Thus the momentum spectrum of the meson is described as
\begin{equation}
\label{production}
\frac{dN_{M}}{d^2p}=\frac{d(N_{M}^{\mathcal{TT}}+N_{M}^{\mathcal{TS}}
+N_{M}^{\mathcal{SS}(1)}+N_{M}^{\mathcal{SS}(2)})}{d^2p}.
\end{equation}
It has been demonstrated in Ref.\cite{CBYpion} that contribution coming from $\mathcal{SS}(2)$ is negligible at RHIC because of the small density of jet and the small overlap probability of the different jets. So $\mathcal{SS}(2)$ term is not considered in the calculation at RHIC, e.g. $\mathcal{SS}=\mathcal{SS}(1)$.

In Refs.\cite{pengru1,pengru2} we have calculated $J/\psi$ transverse momentum spectra for different centralities in Au+Au collisions at $\sqrt{s_{NN}}=200$ GeV at RHIC and the results fit the experimental data well. The contributions coming from $\mathcal{TT}$ and $\mathcal{TS}$ become lower than that of $\mathcal{SS}$ for $p_T>5.8$ GeV/$c$ \cite{pengru1}. The details of calculating $\mathcal{TT}$ and $\mathcal{TS}$ terms are given in Refs.\cite{pengru1,pengru2,pengru3} where the thermal parton distribution is determined by fitting the low-$p_T$ data of $J/\psi$. There are two parameters fitted by the experimental data at RHIC, fugacity of charm quark $\gamma_c=0.26$ and the flow velocity $v_T=0.3c$. And we will still use these two parameters since there are not information of $J/\psi$ production at LHC so far. Now we focus our attention on $\mathcal{SS}$ component which is the dominant contribution to the charmed meson production in the region of $10<p_T<20$ GeV/$c$ discussed in this paper.

The basal idea of the recombination model is to describe the fragmentation process as the recombination of shower partons created in a jet \cite{CBYShower}. The fragmentation process of a parton $i$ splitting into a meson $M$
is expressed as the fragmentation function, then the recombination of two shower partons from one jet $\mathcal{SS}(1)$ is related to FF $D^{M}_{i}$ \cite{CBYpion,pengru1}.With the abbreviation $p$ for the meson transverse momentum, the corresponding one-jet contribution to the inclusive distribution is
\begin{equation}
\label{SS1}
\frac{dN^{\mathcal{SS}(1)}_{M}}{pdpd\phi}=\frac{1}{p^{0}p}\sum_{i}\int\frac{dq}{q}F_{i}(q,\phi,c)
\frac{p}{q}D^{M}_{i}(\frac{p}{q}),
\end{equation}
where fragmentation function $D^{M}_{i}$ describes the possibility of parton $i$ splitting into a meson $M$. FFs for $J/\psi$ and other $D$ mesons can be found in Refs.\cite{Fraggpsi,Fragcpsi} and Refs.\cite{pengru1,pengru2,FragD0}, respectively, including $D^{J/\psi}_g$, $D^{J/\psi}_c$, $D^{D^0}_g$, $D^{D^0}_c$, $D^{D_s}_g$ and $D^{D_s}_c$.
$F_{i}(q,\phi,c)$ is the probability of a hard parton $i$ with momentum $q$ at azimuthal angle $\phi$ in a heavy-ion collision with centrality $c$. For a hard parton created with the distribution $f_i(k)$ at the creation point, the initial momentum $k$ changes into $q$ after traversing an absorptive distance $\xi$. The corresponding distribution is expressed by the momentum degradation factor $G(k,q,\xi)$
\begin{equation}
F_i(q,\xi)=\int dkkf_i(k)G(k,q,\xi).
\end{equation}
In terms of $\xi$, $G(k,q,\xi)$ can be written as a simple exponential form $G(k,q,\xi)=q\delta(q-ke^{-\xi})$ \cite{Yang2010}. If $P(\xi,\phi,c)$ is the probability of having a dynamical path length $\xi$ for a parton directed at $\phi$, the hard parton distribution is obtained after carrying out the integration over $\xi$
\begin{equation}
\label{phidependence}
F_{i}(q,\phi,c)=\int d\xi P(\xi,\phi,c)F_{i}(q,\xi).
\end{equation}
In Ref.\cite{Yang2010}, a scaling behavior of $P(\xi,\phi,c)$ is found for the dependencies on $\phi$ and $c$ for pion production. $P(\xi,\phi,c)$ can be written as a universal function in terms of a scaling variable $z=\xi/\overline{\xi}(\phi,c)$
\begin{equation}
P(\xi,\phi,c)=\psi(z)/\overline{\xi}(\phi,c),
\end{equation}
where $\overline{\xi}$ is the mean dynamical length and the scaling function $\psi(z)$ has been given in Ref.\cite{Yang2010} which will be used in our calculations. In this paper we discuss $J/\psi$ production in central Pb+Pb collisions at LHC with centrality $c=0.05$ (0-10\% centrality) and the initial transverse momentum spectra $f_i(k)$ of hard partons at midrapidity at LHC are parametrized in Refs.\cite{hardparton}.

The contribution which involves shower partons from two different jets is negligible at RHIC due to the small overlap probability for two neighboring jets \cite{CBYpion}. At higher energy, such as LHC, because of the high jet density the new component of the two-jet contribution $\mathcal{SS}(2)$ should be taken into account, which is given by
\begin{eqnarray}
\label{SS2}
\frac{dN^{\mathcal{SS}(2)}}{pdpd\phi}&=&\frac{1}{p^0p}\sum_{i,i'}\int\frac{dq}{q}\frac{dq'}{q'}
F_{i}(q,\phi,c)F_{i'}(q',\phi,c)\Gamma(q,q')\nonumber\\
&  &\times\int\frac{dp_1}{p_1}\frac{dp_2}{p_2}F_{ii'}(q,q';p_1,p_2)R_M(p_1,p_2,p).
\end{eqnarray}
In the above expression, $F_{ii'}$ is the distribution of shower partons related to the two jets and for a meson it is written as
\begin{equation}
F_{ii'}(q,q';p_1,p_2)=S^j_i(\frac{p_1}{q})S^{j'}_{i'}(\frac{p_2}{q'}),
\end{equation}
and $S^j_i(z)$ is the SPD of shower parton $j$ with momentum fraction $z$ in a jet initiated by hard parton $i$.
Another fundamental concept of the recombination model is that the recombination process involves quarks and antiquarks, while gluons are converted to quark-antiquark pairs in the sea before hadronization \cite{Hwa}. Our view is that a hard parton creates a shower of partons that recombine subsequently to form hadrons. Thus gluons are not included in the shower partons, e.g., $i=u(\overline{u}),d(\overline{d}),s(\overline{s}),c(\overline{c}),g$ and $j=u(\overline{u}),d(\overline{d}),s(\overline{s}),c(\overline{c})$.
$R_M(p_1,p_2,p)$ is the recombination function (RF) for the process of the two constituent quarks with momentum $p_1$ and $p_2$ to form a meson with $p$. The parameterized results of the SPDs and RF for the charmed mesons have been discussed before in Refs.\cite{pengru1,pengru2}. The only function we need to know is the overlap function $\Gamma(q,q')$ between the two neighboring jets, which reflects the probability for the overlap of the two shower partons. It depends on the density of jet, or on the energy $\sqrt{s_{NN}}$, the momentum vectors $\vec{q}$ and $\vec{q'}$ of the hard partons, and the width of their jet cones. Since we have not sufficient information of such dependencies for collisions at LHC, the overlap function is approximated by an average quantity $\Gamma$ which varies over a wide range $\Gamma=10^{-n}$ with $n=1, 2,3$ and 4.

\begin{figure}[h]
   \centering
   \includegraphics{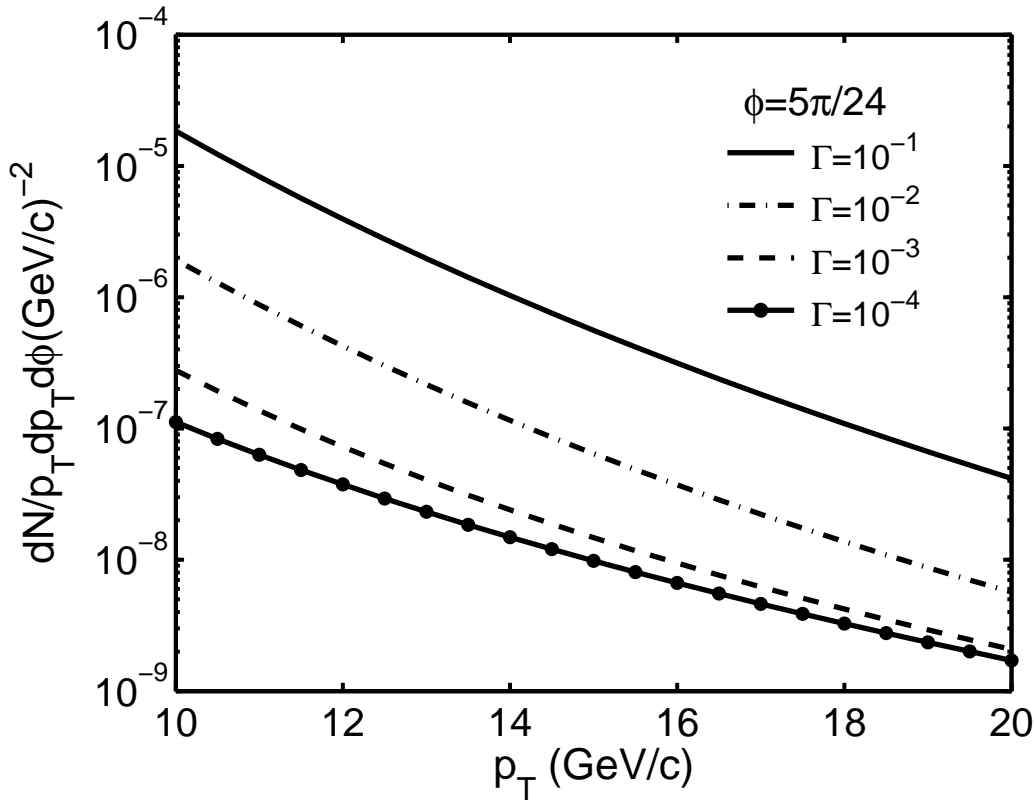}
    \caption{$J/\psi$ distribution as functions of the transverse momentum $p_T$ for the four different values of $\Gamma$.}
  \label{Jpsiptdist}
\end{figure}

\begin{figure}[h]
   \centering
   \includegraphics{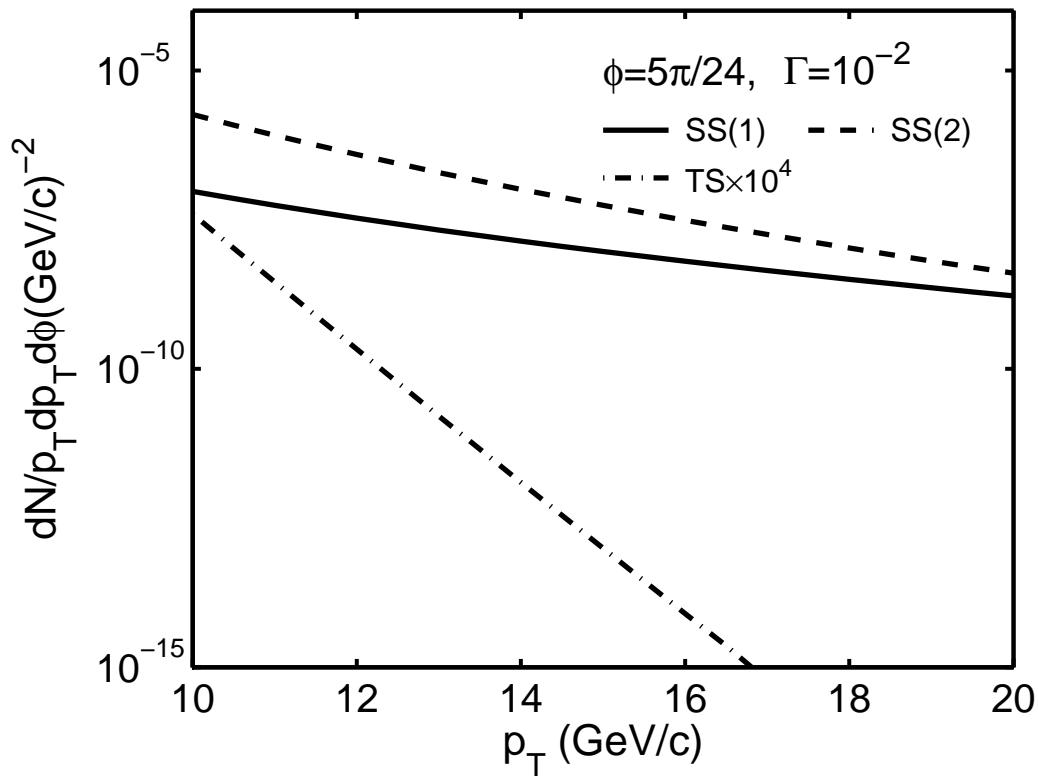}
    \caption{The comparison of three terms for $J/\psi$ with $\Gamma=10^{-2}$.}
  \label{ptdist3}
\end{figure}

The results of the inclusive $J/\psi$ distribution at a given azimuthal angle $\phi=5\pi/24$ for four values of $\Gamma$ are shown in Fig.\ref{Jpsiptdist}. In order to take a clear comparison of the contribution from each term we illustrate them (with $\Gamma=10^{-2}$ for $\mathcal{SS}(2)$) in Fig.\ref{ptdist3} where $\mathcal{TT}$ term is too small to be shown in the figure. As $\Gamma$ decreases, the corresponding probability of the overlap between two neighboring jet cones becomes small and the distributions get close automatically, as shown in Fig.\ref{Jpsiptdist}. In the $p_T$ range shown, the $\mathcal{TS}$ term is about four orders magnitude lower than $\mathcal{SS}(1)$ and  $\mathcal{SS}(2)$ because of the rapidly decreasing thermal parton distribution. The contribution coming from $\mathcal{SS}(2)$ decreases with $p_T$ faster than that from $\mathcal{SS}(1)$ and $\mathcal{SS}(1)$ exceeds over $\mathcal{SS}(2)$ when $\Gamma$ is smaller than $10^{-3}$. Thus the difference between the distributions of the four values of $\Gamma$ decreases with increasing $p_T$.

\begin{figure}[h]
   \centering
   \includegraphics{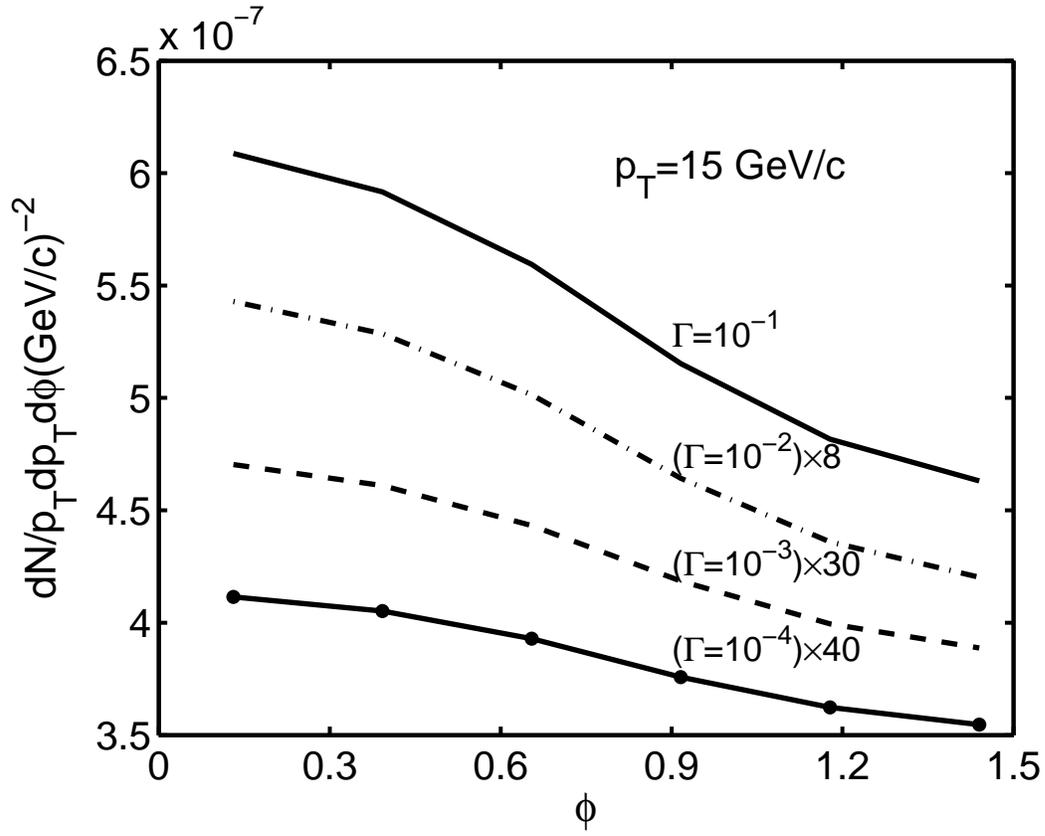}
    \caption{The azimuthal anisotropy of $J\psi$ distribution for the four different values of $\Gamma$.}
  \label{Jpsiphidist}
\end{figure}

\begin{figure}[h]
   \centering
   \includegraphics{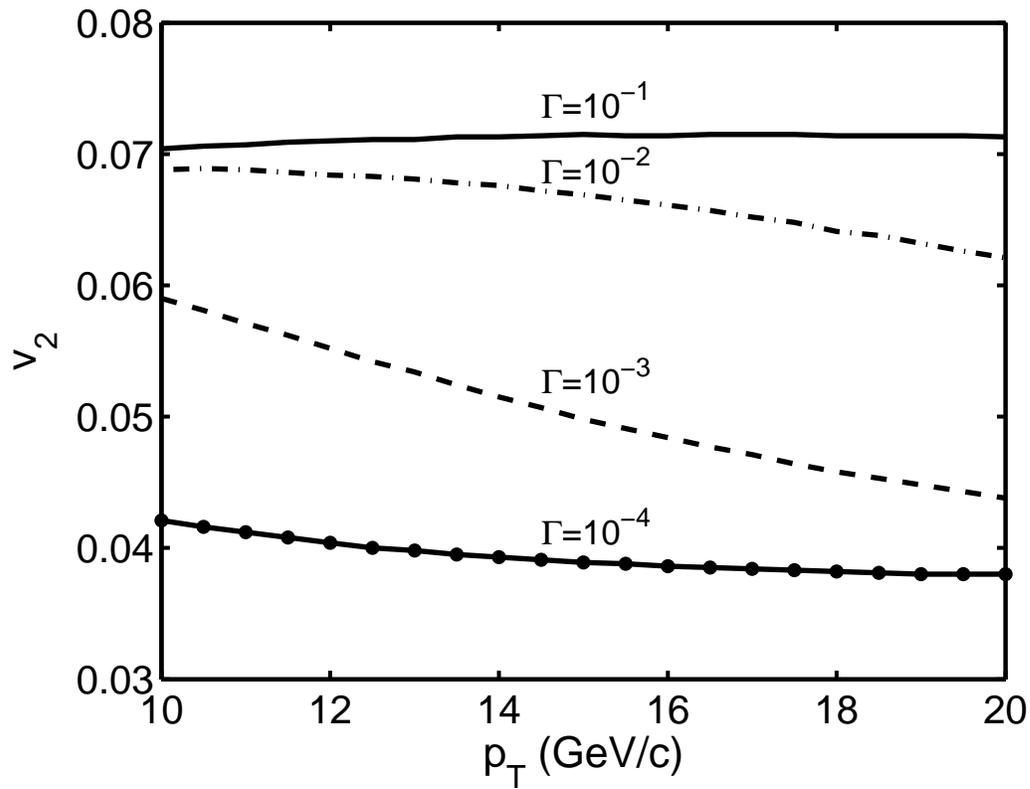}
    \caption{The predicted $v_2$ of $J/\psi$ for the four different values of $\Gamma$.}
  \label{Jpsiv2}
\end{figure}

In Fig.\ref{Jpsiphidist}, we show the azimuthal anisotropy of the transverse momentum distribution of $J/\psi$ at given $p_T=15$ GeV/c as a function of the azimuthal angle. We can get the elliptic flow $v_2$ for different values of $\Gamma$
for such a $\phi$ dependence through
\begin{equation}
\frac{dN}{p_Tdp_Td\phi}=A(p_T)[1+2v_2(p_T)\cos(2\phi)].
\end{equation}
Shown in Fig.\ref{Jpsiv2} is the predicted $v_2$ of $J/\psi$ as a function of $p_T$ for different $\Gamma$. The $v_2$ value increases with $\Gamma$. And the values of $v_2$ for $\Gamma$ from $10^{-3}$ to $10^{-1}$ are larger than that at RHIC calculated in Ref.\cite{pengru3}, where $v_2$ increases monotonically up to $p_T\simeq4$ GeV/c and then saturates at 0.04 for 0-10\% centrality. If $\Gamma$ is small enough ($\Gamma=10^{-4}$) so that the impact of $\mathcal{SS}(2)$ can be ignored, the value of $v_2$ maintains as a constant about 0.04 which is almost the same as the saturated constant calculated at RHIC \cite{pengru3}. When $\Gamma$ increases to $10^{-1}$ the $\mathcal{SS}(2)$ term becomes so large that the other three terms are negligible and $v_2$ also keeps as a constant of 0.07.
The main reason is that when $\Gamma$ is large the contribution from $\mathcal{SS}(2)$ plays an important role because there are two terms contributing to the azimuthal angle dependence in $\mathcal{SS}$ recombination.

The results of $v_2$ for $\Gamma=10^{-3}$ and $\Gamma=10^{-2}$ decrease with increasing $p_T$. In fact
the trend that $v_2$ decreases with $p_T$ has been probed in heavy ion collisions at RHIC at $p_T>3$ GeV/$c$ \cite{STARv2,EXPv2decrease} (review in Ref.\cite{reviewEXP}). STAR Collaboration presented the results of the azimuthal anisotropy parameter $v_2$ for Au+Au collisions at $\sqrt{S_{NN}}=200$ GeV extend the measurements up to $p_T=12$ GeV/$c$ \cite{STARv22} from the reaction plane method and four-particle cumulant method. It shows that $v_2$ is maximum at $p_T\sim3$ GeV/$c$, saturates at $3<p_T<6$ GeV/$c$ and then decreases slowly up to the highest momentum  measured. Thus the reliable discussion of the trends in the region of $p_T>6$ GeV/$c$ requires multi-particle correlation measurements \cite{STARv23}. The eventual decrease of $v_2$ at high $p_T$ has been demonstrated as a consequence of finite inelastic jet energy loss by the perturbative QCD theory \cite{v2decrease}. Another recombination model which describes hadronization
via quark coalescence has also been applied to explain the elliptic flow of mesons and baryons \cite{Coalescence}.

\begin{figure}[h]
   \centering
   \includegraphics{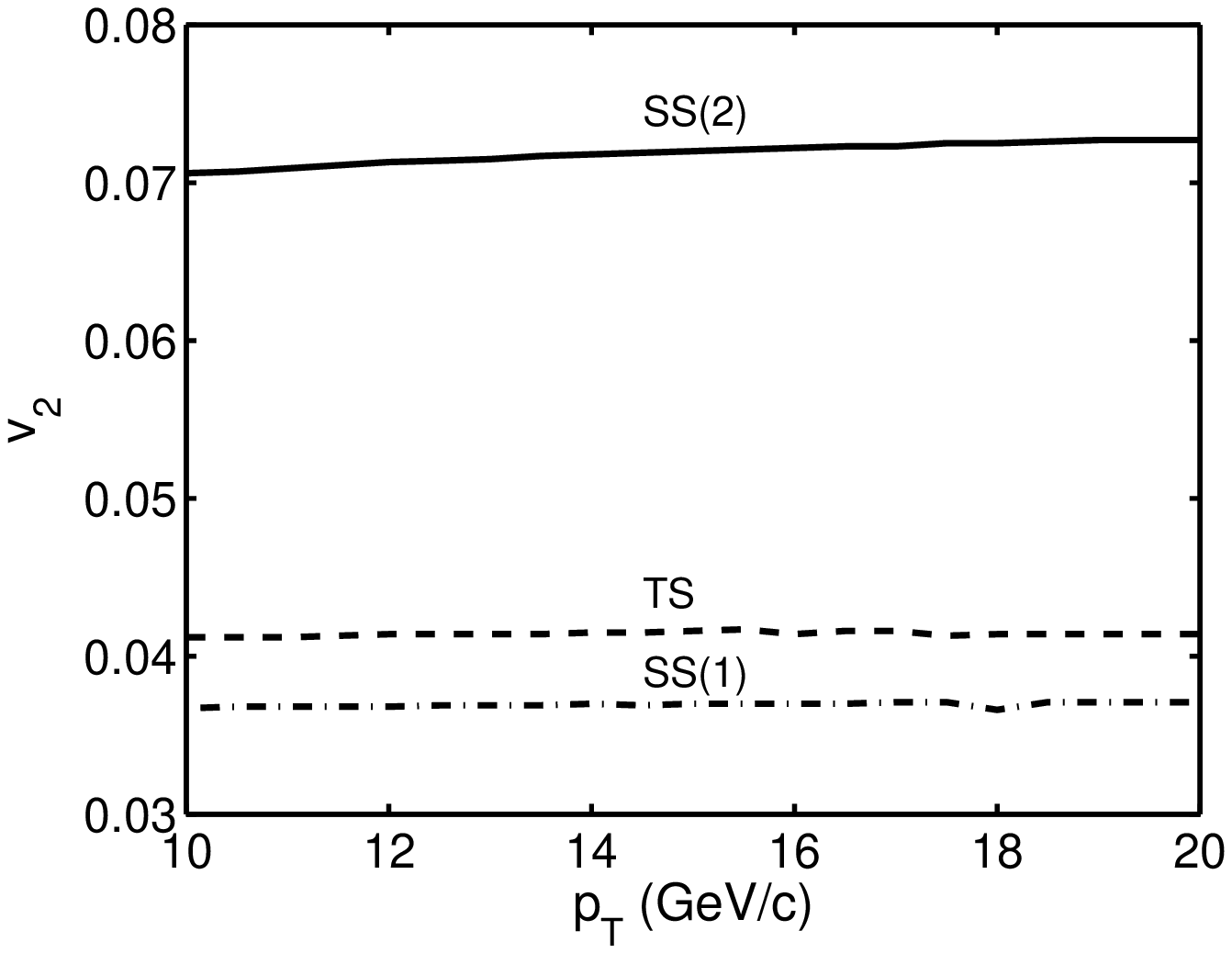}
    \caption{$v_2$ from $\mathcal{TS}$, $\mathcal{SS}(1)$ and $\mathcal{SS}(2)$ terms, respectively.}
  \label{comparev2}
\end{figure}

In Fig.\ref{comparev2}, we calculate $v_2$ coming from the term of $\mathcal{TS}$, $\mathcal{SS}(1)$ and $\mathcal{SS}(2)$, respectively. The values of $v_2$ from $\mathcal{TS}$ and $\mathcal{SS}(1)$ terms are almost constants of 0.041 and 0.038. While $v_2$ from $\mathcal{SS}(2)$ increases with $p_T$ slightly and at the same time it is much larger than those from $\mathcal{TS}$ and $\mathcal{SS}(1)$. It's a reasonable result. If the meson is formed by the recombination of thermal-shower partons or shower-shower partons from the same jet, the process is related to only one jet. While two different jets should be considered in the recombination of $\mathcal{SS}(2)$. Shower partons from two jets may undergo more energy loss than from one jet before they are formed into a meson and the energy loss depends on the azimuthal angle $\phi$ of the hard parton in Eq.(\ref{phidependence}). Thus the value of $v_2$ associated to two different jets is larger than the others. Also we have discussed that the contribution coming from $\mathcal{SS}(2)$ decreases faster than that from $\mathcal{SS}(1)$ when $p_T$ increases. So when the contributions to the flow from the terms of $\mathcal{SS}(1)$ and $\mathcal{SS}(2)$ are comparative ($\Gamma=10^{-3}, 10^{-2}$), the $\phi$ dependence of the distribution of $J/\psi$ will decrease with increasing $p_T$, as shown in Fig.\ref{Jpsiv2}.

\begin{figure}[h]
   \centering
   \includegraphics{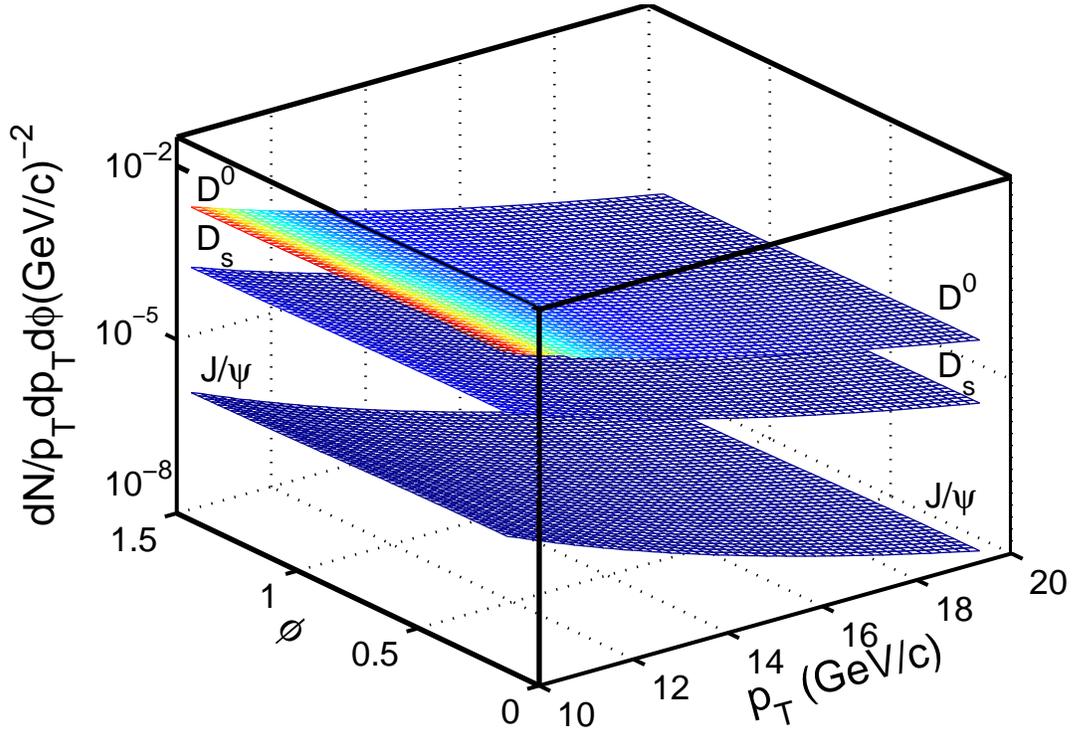}
    \caption{The distribution of $J/\psi$, $D^0$ and $D_s$ as functions of the transverse momentum $p_T$ and the azimuthal angle $\phi$ with $\Gamma=10^{-2}$.}
  \label{ptphidist}
\end{figure}

Finally we predict the production of the other two charmed mesons $D^0$ and $D_s$ and the dependence on the transverse momentum and the azimuthal angle is illustrated in Fig.\ref{ptphidist} with $\Gamma=10^{-2}$.

In summary, we predict the distribution of $J/\psi$ and its elliptic flow parameter $v_2$ at LHC energy in the framework of the recombination model. Because of the high energy density the recombination of shower partons from two neighboring jets may play an important role in the meson production and its influence to $v_2$ is discussed. And $v_2$ of $J/\psi$ is predicted to decrease with $p_T$.

This work was supported in part by the National Natural Science Foundation of China under Grant Nos. 10635020 and 10775057, by the Ministry of Education of China under Grant No. 306022 and project IRT0624, and by the Programme of Introducing Talents of Discipline to Universities under No. B08033.

\end{document}